\documentstyle [12pt]{article}
\makeatletter \oddsidemargin  0in \evensidemargin 0in
\textwidth16cm \RequirePackage[dvips]{graphicx} \textheight 20.5cm
\newcommand{\singlespacing}{\let\CS=\@currsize\renewcommand{\baselinestretch}{1.5}\tiny\CS}
\makeatletter \oddsidemargin  -.1in \evensidemargin -.1in
\newcommand{\doublespacing}{\let\CS=\@currsize\renewcommand{\baselinestretch}{1.35}\tiny\CS}

\def\@citex[#1]#2{\if@filesw\immediate\write\@auxout{\string\citation{#2}}\fi
  \def\@citea{}\@cite{\@for\@citeb:=#2\do
    {\@citea\def\@citea{,\linebreak[0]\hskip0pt plus .2em}%
      \@ifundefined{b@\@citeb}%
    {{\bf ?}\@warning{Citation `\@citeb' on page \thepage\space undefined}}%
      \hbox{\csname b@\@citeb\endcsname}}}{#1}}

\newtheorem{rule-def}[theorem]{Rule}
\begin{document}
\title{\bf Self Replication and Signalling}
\author{I.Chakrabarty$^{a}$ \thanks{Corresponding author:
E-Mail-indranilc@indiainfo.com }, Prashant$^{b}$, B.S.Choudhury $^{c}$\\
$^{a}$ Department of Mathematics,\\ Heritage Institute of
Technology,Kolkata-107,West Bengal,India\\ $^{b}$ Indian Institute of Information Technology and Management, Gwalior,MP, India\\
$^{c}$Department of Mathematics \\ Bengal Engineering and Science
University, Howrah, India}
\date{}
\maketitle{}
\begin{abstract}
It is known that if one could clone an arbitrary quantum state
one could send signal faster than the speed of light. However it
remains interesting to see that if one can perfectly self
replicate an arbitrary quantum state, does it violate the no
signalling principle? Here we see that perfect self replication
would also lead to superluminal signalling.
\end{abstract}
\section{\bf Introduction}
It was Shannon who first introduced us to the amazing world of 0
and 1. After that it had been a long journey and still we are
searching for the nature of information in quantum world.  One of
the major thrust area of research is to look out how the
information defined in classical world differs from quantum
information and also to enlist many computational procedures which
are feasible in the classical world but get restricted at
microscopic level. Cloning and deletion which is feasible in
classical information, cannot be executed with hundred percent
fidelity for an unknown quantum state [1,2]. Recently it had been
seen that self replication machine which is feasible in classical
information theory [3] fails to self replicate two non orthogonal
quantum states [4]. It may be remarked that the process of
replication or cloning and self replication is not just the same
thing. For the proper understanding of the matter
we are introducing the concept of universal constructor.\\
{\bf A Universal Constructor}: A Universal constructor or a
quantum mechanical self replicating machine is defined as a
machine which can implement the copying of the original state
along with the stored programme by a linear operator acting on the
joint Hilbert Space. This machine may be completely specified by a
quadruple $(|\psi\rangle,|P_{U}\rangle,|C\rangle,|\Sigma\rangle)$.
Here $|\psi\rangle \in H^{N}$ (where $H$ is the associated Hilbert
space with dimension 'N') is the input state that contains quantum
information to be self replicated and $|P_{U}\rangle \in H^{K}$
(where $K$ is the dimension of the associated Hilbert space)is the
programme state that carries the instructions to copy the original
information (the unitary operator $U$;
$U(|\psi\rangle|0\rangle)=|\psi\rangle|\psi\rangle$, is encoded in
the programme state $|P_{U}\rangle$). Here $|C\rangle$ is the
control qubit, which separates the original information from the
program states so that there remains no quantum entanglement
between the replicated and the original information.
$|\Sigma\rangle$ is the finite collection of n blank states
$|0\rangle|0\rangle|0\rangle.....|0\rangle$ in a $M$ dimensional
Hilbert space, where each $|0\rangle\in H^{N}$. Hence $M=N^{n}$.
In order to copy the program states say m blank states are needed,
so $K=N^{m}$ We had already seen in [4,5] that self replication of
two non-orthogonal programme state is not feasible in general.
However for self replication of a finite number K of the
non-orthogonal states with K the dimension of the program Hilbert
space, then it may be possible to design a universal
constructor with finite resources.\\
In this work we try to see that whether the principle of no self
replication is consistent with the principle of no signalling. The
principle of no signalling states that : If two parties (say Alice
and Bob) are separated by space like distance, then one cannot
send superluminal signal faster than the speed of light. All the
no go theorems like cloning , deletion , flipping are consistent
with this principle of no signalling [6,7,8]. Here we will see
that 'no-self replication' theorem is also consistent with the
principle of no signalling. In this problem we will consider a
hypothetical situation where two distant parties Alice and Bob
share an entangled state. Now if Alice  self replicates her qubit
with the help of a universal constructor, then Alice can
distinguish two statistical mixture representing her subsystem as
a consequence of Bob's measurement on two different qubit basis.
As a result of which Alice would tell on which basis Bob has
measured his qubit. This will imply superluminal signalling.
Interestingly in the next section we are going to show that this
is not going to be true.
\section {\bf Principle of no signalling and self replication :}
Let us consider two entangled states shared by two distant
partners Alice and Bob. The entangled states are given as
\begin{eqnarray}
|\chi_1\rangle=\frac{1}{\sqrt{2}}[|\psi_1\rangle_A|\psi_2\rangle_B-|\psi_2\rangle_A|\psi_1\rangle_B]\nonumber\\
|\chi_2\rangle=\frac{1}{\sqrt{2}}[|P_{u1}\rangle_A|P_{u2}\rangle_B-|P_{u2}\rangle_A|P_{u1}\rangle_B]
\end{eqnarray}
where $A$ denotes that the qubit is in possession with Alice, and
$B$ denotes Bob's qubit. The states
$|\psi_1\rangle,|\psi_2\rangle$ are two non orthogonal quantum
states whereas the states $|P_{u1}\rangle,|P_{u2}\rangle$ are non
orthogonal quantum states. The combined state of Alice and bob is
given by,
\begin{eqnarray}
|\chi_{12}\rangle&=&\frac{1}{2}[(|\psi_1\rangle|P_{u1}\rangle)_{A}(|\psi_2\rangle|P_{u2}\rangle)_{B}-
(|\psi_2\rangle|P_{u1}\rangle)_{A}(|\psi_1\rangle|P_{u2}\rangle)_{B}-
{}\nonumber\\&&(|\psi_1\rangle|P_{u2}\rangle)_A(|\psi_2\rangle|P_{u1}\rangle)_B-
(|\psi_2\rangle|P_{u2}\rangle)_{A}(|\psi_1\rangle|P_{u1}\rangle)_{B}]
\end{eqnarray}
 The principle of no signalling tells
us that if some local operation is done on Bob's qubit ,it will
not change the density matrix describing Alice's subsystem. Now ,
in this context , where Alice and Bob are sharing entangled states
(2), if we trace out Bob's qubit, the density matrix describing
Alice's subsystem is given by,
\begin{eqnarray}
\rho_A&=&\frac{1}{4}[|\psi_1P_{u1}\rangle\langle
\psi_1P_{u1}|+|\psi_2P_{u1}\rangle\langle
\psi_2P_{u1}|+|\psi_1P_{u2}\rangle\langle
\psi_1P_{u2}|+|\psi_2P_{u2}\rangle\langle
\psi_2P_{u2}|{}\nonumber\\&&-|\psi_2P_{u1}\rangle\langle
\psi_1P_{u1}|(\langle
\psi_2|\psi_1\rangle)-|\psi_1P_{u2}\rangle\langle
\psi_1P_{u1}|(\langle
P_{u2}|P_{u1}\rangle){}\nonumber\\&&+|\psi_2P_{u2}\rangle\langle
\psi_1P_{u1}|(\langle \psi_2|\psi_1\rangle\langle
P_{u2}|P_{u1}\rangle)-|\psi_1P_{u1}\rangle\langle
\psi_2P_{u1}|(\langle
\psi_1|\psi_2\rangle){}\nonumber\\&&+|\psi_1P_{u2}\rangle\langle
\psi_2P_{u1}|(\langle \psi_1|\psi_2\rangle\langle P_{u2}
|P_{u1}\rangle)-|\psi_2P_{u2}\rangle\langle \psi_2P_{u1}|(\langle
P_{u2}|P_{u1}\rangle){}\nonumber\\&&-|\psi_1P_{u1}\rangle\langle
\psi_1P_{u2}|(\langle
P_{u1}|P_{u2}\rangle)+|\psi_2P_{u1}\rangle\langle
\psi_1P_{u2}|(\langle \psi_2|\psi_1\rangle\langle
P_{u1}|P_{u2}\rangle){}\nonumber\\&&-|\psi_2P_{u2}\rangle\langle
\psi_1P_{u2}|(\langle
\psi_1|\psi_2\rangle)+|\psi_1P_{u1}\rangle\langle
\psi_2P_{u2}|(\langle \psi_1|\psi_2\rangle\langle
P_{u1}|P_{u2}\rangle){}\nonumber\\&&-|\psi_2P_{u1}\rangle\langle
\psi_2P_{u2}|(\langle
P_{u1}|P_{u2}\rangle)-|\psi_1P_{u2}\rangle\langle
\psi_2P_{u2}|(\langle \psi_1|\psi_2\rangle)]
\end{eqnarray}
At this point one can ask an interesting question that if Bob can
self replicate his qubit, then will there be any change in the
density matrix (3) describing Alice's subsystem. The principle of
no signalling tells us that it is impossible to achieve.\\
Let us Bob possesses a  hypothetical quantum mechanical universal
constructor, which will self replicate the qubit possessed by Bob.
The action of the self replicating machine is given by,
\begin{eqnarray}
&&|\psi_1\rangle|0\rangle|P_{u_1}\rangle|0\rangle^{\otimes
m}|C\rangle|0\rangle^{\otimes
n-(m+1)}\longrightarrow|\psi_1\rangle|P_{u_1}\rangle(|\psi_1\rangle|0\rangle|P_{u_1}\rangle|0\rangle^{\otimes
m}|C_1\rangle)|0\rangle^{\otimes n-2(m+1)}{}\nonumber\\&&
|\psi_2\rangle|0\rangle|P_{u_1}\rangle|0\rangle^{\otimes
m}|C\rangle|0\rangle^{\otimes
n-(m+1)}\longrightarrow|\Phi_1\rangle{}\nonumber\\&&
|\psi_1\rangle|0\rangle|P_{u_2}\rangle|0\rangle^{\otimes
m}|C\rangle|0\rangle^{\otimes
n-(m+1)}\longrightarrow|\Phi_2\rangle{}\nonumber\\&&
|\psi_2\rangle|0\rangle|P_{u_2}\rangle|0\rangle^{\otimes
m}|C\rangle|0\rangle^{\otimes
n-(m+1)}\longrightarrow|\psi_2\rangle|P_{u_2}\rangle(|\psi_2\rangle|0\rangle|P_{u_2}\rangle|0\rangle^{\otimes
m}|C_2\rangle)|0\rangle^{\otimes n-2(m+1)}
\end{eqnarray}
Here $|P_{u_1}\rangle$ and $|P_{u_2}\rangle$ are the program
states where the instruction to copy the original information has
been encoded. Here the state $|C\rangle$ acts as an control unit
which separates the original with the replica in such a manner so
that there exists no quantum entanglement between them. Here
$|C_1\rangle$ and $|C_2\rangle$ are the respective states of the
control unit at the output port. The states $|\Phi_1\rangle$ and
$|\Phi_2\rangle$ are the composite states at the output port.\\
It is important to note that the above transformations (2-3)are
not merely a cloning transformation, on the contrary it is a
recursively defined transformation where the fixed unitary
operator acts on initial configuration and the same operator acts
on the final configuration after the copies have been produced.\\
Now Bob attaches ancilla states
$|\Sigma\rangle=|0\rangle^{\otimes n}$as well as the control
qubit $|C\rangle$ to his subsystem. Now Bob applies the
transformations defined in (3) on his qubit, in such a way that
out of these n blank states he uses only one blank state for the
cloning of the original states, m blank states for the cloning of
the programmed states and he is left with $n-(m+1)$ blank states
for future replication, until all the n blank states are
exhausted. He also uses the control qubit to ensure that there
remains no entanglement between the original and replica. As
result of his action the entangled state shared between these two
parties Alice and Bob takes the form,
\begin{eqnarray}
|\chi_{12}^S\rangle=\frac{1}{2}[(|\psi_1\rangle|P_{u1}\rangle)|Y\rangle-(|\psi_2\rangle|P_{u1}\rangle)|\Phi_1\rangle-
(|\psi_1\rangle|P_{u2}\rangle)|\Phi_2\rangle+(|\psi_2\rangle|P_{u2}\rangle)|X\rangle]
\end{eqnarray}
where,\\
$|Y\rangle=|\psi_2\rangle|P_{u_2}\rangle(|\psi_2\rangle|0\rangle|P_{u_2}\rangle|0\rangle^{\otimes
m}|C_2\rangle)|0\rangle^{\otimes
n-2(m+1)},\\|X\rangle=|\psi_1\rangle|P_{u_1}\rangle(|\psi_1\rangle|0\rangle|P_{u_1}\rangle|0\rangle^{\otimes
m}|C_1\rangle)|0\rangle^{\otimes n-2(m+1)}$\\
 Now if one traces out Bob's qubit , then the density matrix
representing Alice's subsystem is given by,
\begin{eqnarray}
\rho_A^{S}&&=\frac{1}{4}[|\psi_1P_{u1}\rangle\langle
\psi_1P_{u1}|+|\psi_1P_{u2}\rangle\langle
\psi_1P_{u2}|+|\psi_2P_{u1}\rangle\langle
\psi_2P_{u1}|+|\psi_2P_{u2}\rangle\langle
\psi_2P_{u2}|{}\nonumber\\&&+|\psi_2P_{u2}\rangle\langle
\psi_1P_{u1}|(\langle Y |X\rangle)-|\psi_2P_{u1}\rangle\langle
\psi_1P_{u1}|(\langle Y
|\Phi_1\rangle)-|\psi_1P_{u2}\rangle\langle \psi_1P_{u1}|(\langle
Y |\Phi_2\rangle){}\nonumber\\&&-|\psi_1P_{u1}\rangle\langle
\psi_2P_{u1}|(\langle \Phi_1
|Y\rangle)+|\psi_1P_{u2}\rangle\langle \psi_2P_{u1}|(\langle
\Phi_1 |\Phi_2\rangle)-|\psi_2P_{u2}\rangle\langle
\psi_2P_{u1}|(\langle \Phi_1
|X\rangle){}\nonumber\\&&-|\psi_1P_{u1}\rangle\langle
\psi_1P_{u2}|(\langle \Phi_2 |Y\rangle)+
|\psi_2P_{u1}\rangle\langle \psi_1P_{u2}|(\langle \Phi_2
|\Phi_1\rangle)-|\psi_2P_{u2}\rangle\langle \psi_1P_{u2}|(\langle
\Phi_2 |X\rangle){}\nonumber\\&&-|\psi_2P_{u1}\rangle\langle
\psi_2P_{u2}|(\langle X
|\Phi_1\rangle)-|\psi_1P_{u2}\rangle\langle \psi_2P_{u2}|(\langle
X |\Phi_2\rangle)+|\psi_1P_{u1}\rangle\langle
\psi_2P_{u2}|(\langle X |Y\rangle)]
\end{eqnarray}
Since the density matrices (3) and (6) representing Alice's
subsystem before and after the application  of Universal self
replicating machine on his qubit by Bob are not identical, we
conclude that signalling has taken place due to local action
performed by Bob. However no signalling principle tells us that
this is not possible in reality. Hence we conclude that perfect
self replication for arbitrary quantum states is not possible from
the principle of no signalling.\\
\section{\bf Conclusion:} In summary we can say that this work
once again shows the peaceful coexistence of no go theorem like no
self replication theorem with the physical principles like
principle of no signalling. In other words, one cannot self
replicate arbitrary quantum states with the help of program state
as this will violate the principle of no signalling.
\section{\bf Acknowledgement}
Indranil Chakrabarty acknowledges S. Adhikari,
Prof.C.G.Chakraborti for their encouragement and inspiration in
completing this work. Prashant acknowledges Almighty for getting
the inspiration for research and also Prof Gilles Bassard for
providing encouragement and support in complementing the work .
Both Indranil and Prashant acknowledges Prof A.K.Pati, for
providing encouragement and support in completing this work.
\section{\bf Reference:}
$[1]$ W.K.Wootters and W.H.Zurek,Nature \textbf{299}, 802 (1982) .\\
$[2]$ A.K.Pati and S.L.Braunstein Nature \textbf{404}, 164 (2000) .\\
$[3]$ J.Von Neumann, The theory of Self-Replicating Automata.
University of Illinois Press,Urbana,IL (1966)(work by J.von
Neumann in 1952-53).\\
$[4]$ A.K.Pati and S.L.Braunstein, Quantum mechanical universal
constructor,quantph /0303124 (2003).\\
$[5]$ M. A. Nielsen, I. Chuang, Phys. Rev. Lett. 79, 321
(1997).\\
$[6]$ A.K.Pati, S.L.Braunstein, Phys. Lett A \textbf{315} , 208 (2003).\\
$[7]$ I.Chattopadhyay.et.al.Phys. Lett. A, \textbf{351}, 384 (2006).\\
$[8]$ A.K.Pati, Phys. Lett. A, \textbf{292}, 12 (2001).
\end{document}